\documentclass[sn-mathphys]{sn-jnl}

\usepackage{amssymb}
\usepackage{graphicx}
\usepackage{epstopdf}
\usepackage{amsmath}
\usepackage{epsfig}
\usepackage{color}
\usepackage{ulem}
\usepackage{stmaryrd}

\jyear{2021}

\theoremstyle{thmstyleone}

\theoremstyle{thmstyletwo}

\theoremstyle{thmstylethree}

\begin{document}

\title[Statistical and Mathematical Evidence of Rigged Parliamentary Elections in Georgia, 2024]{Statistical and Mathematical Evidence of Rigged Parliamentary Elections in Georgia, 2024}

\author*[1]{\fnm{Lazare} \sur{Osmanov}}\email{losma21@freeuni.edu.ge}

\author[2]{\fnm{Levan} \sur{Ghaghanidze}}\email{lghag19@freeuni.edu.ge}

\author[3]{\fnm{Saba} \sur{Sigua}}\email{ssigu2020@agruni.edu.ge}

\author[4]{\fnm{Temur} \sur{Begishvili}}\email{tbegi21@freeuni.edu.ge}

\author[5]{\fnm{Keso} \sur{Bostoghanashvili}}\email{kbost20@freeuni.edu.ge}

\affil*[1]{\orgdiv{School of Physics}, \orgname{Free University of Tbilisi}, \orgaddress{\street{David Aghmashenebeli Alley}, \city{Tbilisi}, \postcode{0159}, \country{Georgia}}}

\affil*[2]{\orgdiv{School of Business and Administration}, \orgname{Free University of Tbilisi}, \orgaddress{\street{David Aghmashenebeli Alley}, \city{Tbilisi}, \postcode{0159}, \country{Georgia}}}

\affil*[3]{\orgdiv{School of Business and Administration}, \orgname{Agricultural University of Georgia}, \orgaddress{\street{David Aghmashenebeli Alley}, \city{Tbilisi}, \postcode{0159}, \country{Georgia}}}

\affil[4]{\orgdiv{school of Physics}, \orgname{Free University Of Tbilisi}, \orgaddress{\street{David Aghmashenebeli Alley}, \city{Tbilisi}, \postcode{0128},  \country{Georgia}}}

\affil[5]{\orgdiv{school of Governance and social sciences}, \orgname{Free University Of Tbilisi}, \orgaddress{\street{David Aghmashenebeli Alley}, \city{Tbilisi}, \postcode{0159},  \country{Georgia}}}

\abstract{The official data provided by "Central Election Commission" was analyzed, revealing irregularities that raised reasonable suspicion of election manipulation by the winning ``Georgian Dream Party." However, these suspicions alone were insufficient to provide concrete evidence. A computational approach was developed based on the official data to address this. Through this analysis, one method estimated the number of manipulated votes to range between 140,000 and 200,000, while another approach estimated a broader range of 90,000 to 245,000 votes. Notably, both methods identified the most probable number of manipulated votes as 175,000, providing a strong mathematical basis to substantiate the claim of election falsification.
}


\maketitle

\section{Introduction}\label{sec1}

On October 26th 2024, the Republic of Georgia had one of the most important elections in history, in which Georgia had to choose between two forces, Pro-European or Pro-Russian. Before this historic election, the Georgian parliament, ruled by the ``Georgian Dream party", proposed a law called ``On transparency of foreign influence". The law regulates Organizations with more than $20\%$ of their income from foreign sources. If an entity is among such organizations, The entity has to register as an organization carrying out the interests of a foreign power. The people of Georgia saw this law as a stigmatizing, hateful law, and aiming to declare Europe and America as enemies, because The largest part of the income comes from Europe and America. The west saw this move by the ruling party as an attempt to turn Georgia into the next Belarus. Also, The same law was passed in Russia in 2012 and the next years in pro-Russian countries. Masses started to protest in the streets of the capital Tbilisi and other major Georgian cities. Despite the people's resistance the ``Georgian Dream Party" proceeded to pass the law. The ongoing situation on the streets of Georgia in spring attracted the attention of MEPs (Members of the European Parliament) and the international media. The Georgian people saw the October election as the only democratic way to change the government. The opposition saw the preelection period as a chance to train the observers, as being an observer was the right way to protect votes. For the first time in history, this election used an electronic voting system leading people to think there would be fewer means to manipulate the votes. In Georgia the entity responsible for conducting elections is the Central election commission (CEC) which divides into District commission. Each district commission combines several election precincts.  Before the election, Edison Research, Gorbi, and Harrisx which are qualitative and quantitative research companies, conducted research and published exit polls. Before $26$'th of October $2024$, exit polls by these three research groups were quite different from one another. Edison Research has been conducting research on exit polls in Georgia since $2012$. On October 26th at 8 p.m., an exit poll made by Edison Research showed that ``The Georgian Dream party's" vote was determined to be $41\%$\cite{a4}, greatly differing from the final result of the election where the votes for the "The Georgian Dream party" totaled to $54\%$ \cite{a4}. This was the only case in history where Edison research's exit polls and the CEC final results presented such a stark difference, with the CEC final results showing a $13\%$ higher number of votes for the incumbent party. In $2012$ Edison Research's exit polls predicted that the "Georgian Dream party" would have $51\%$ of the votes and in the final results they had $54\%$, a similar situation occurred in $2016$, and $2020$. In 2016 the Edison Research exit poll resulted in about $46\%$\cite{a4} for "The Georgian Dream Party" and in reality, they received $48\%$\cite{a4}. In the 2018 presidential election, the Edison Research exit poll showed that the "Georgian Dream party" would receive $40\%$, and in reality, they received $39\%$\cite{a4}.
Therefore, we have decided to conduct a mathematical investigation of the current elections, using computer simulations that are as precise as possible, using only official data from "CEC." Here, we simulate election results such that the overall percentage of votes for "The Georgian Dream Party" is the same as the official election outcome and then observe the simulated distribution of percentages across the different election precincts, comparing them to the official percentage distributions to determine any discrepancies.
The initial section offers the introduction and the analysis of the official data for every party that crossed the $5\%$ threshold. In the following section, we discuss how the code for the election simulation was made and also the analysis of the results. In the next section, we will discuss the results and the number of manipulated votes. Finally, we will sum up our results and conclude that there is evidence that the integrity of the  parliamentary elections of $2024$ in Georgia was undermined. 

\section{Analysis of the official data} \label{sec2}

In this section, we will analyze the data and provide exact information on each graph. In Figure 1, we show the distribution of the percentages of votes received by each of the parties that obtained an overall vote of $5\%$ or more across precincts based on "CEC" official data. The x-axis of the histogram represents the percentage of votes obtained in each precinct, divided into bins of size $5\%$ in panel (a) and bins of size $1\%$ in panel (b). The y-axis corresponds to the number of precincts falling within each percentage bin. For example, in panel (a) ``Georgian Dream Party" (light blue) received from $0\%$-$20\%$ in a small number of precincts, about $40\%$ in 60 precincts, $80\%$ in 25 precincts, and $100\%$ in zero precincts.

\begin{figure}
\resizebox{\hsize}{!}{\includegraphics[width=10cm]{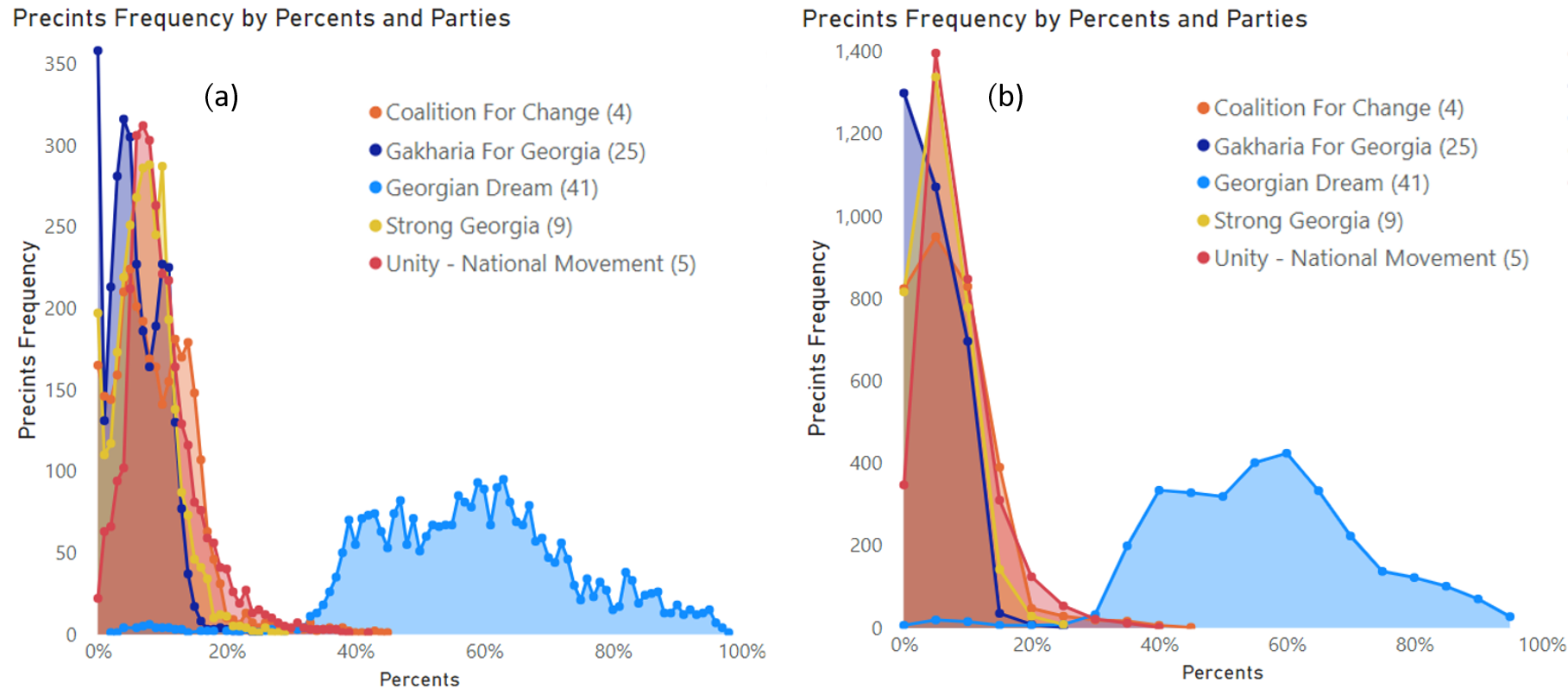}}
\caption{Panel (a) represents a histogram of the precincts based on the percentage of votes received for each of the parties that received an overall vote of $5\%$ or more (Light blue: Georgian Dream Party, orange: Coalition for Change, dark blue:  Gakharia for Georgia, yellow: Strong Georgia, red: Unity - National Movement). The bin size is $5\%$. (b) presents the same distribution with a finer bin size of $1\%$. All the information on both graphs are taken from official "CEC" datasets.} \label{fig1}
\end{figure}

\begin{figure}
\resizebox{\hsize}{!}{\includegraphics[width=10cm]{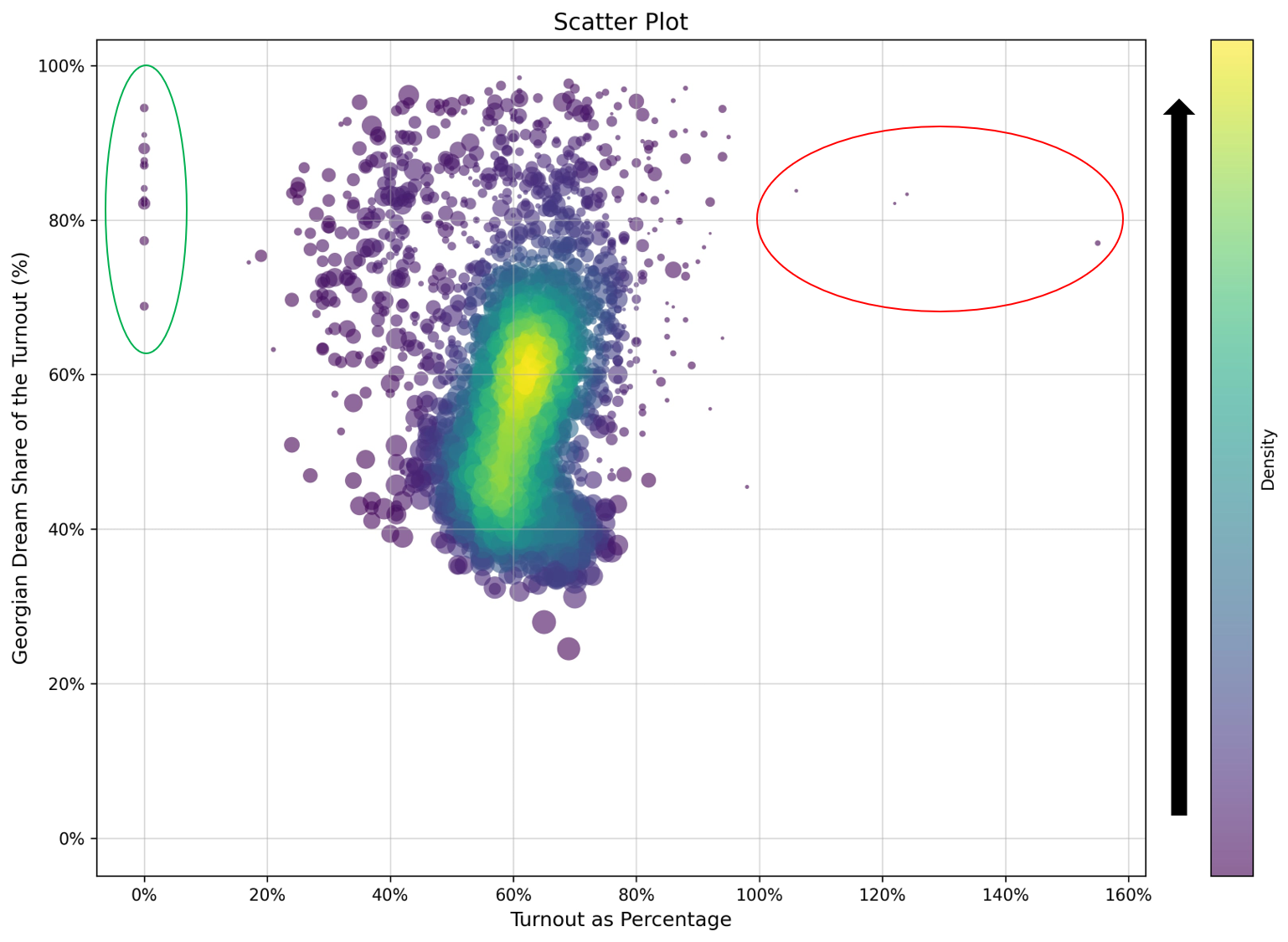}}
\caption{Scatter plot of the voter turnout in each precinct and the share of votes received by ``The Georgian Dream Party". Each point on the graph represents one of the election precincts, the size of the circle indicates the total turnout at the election precinct. The color represents the density of precincts, the brighter the color, the more dense. The red ellipse indicates the precincts in which the turnout exceeded $100\%$. the green ellipse indicates precincts for which there was no information about the number of registered voters.} \label{fig2}
\end{figure}

The ``Coalition for Change" (orange), received $0\%$ of the votes in 160 precincts, $20\%$ in less than 50 precincts, and so on. The same logic applies to the different political parties, and as is shown in the figure ``The Georgian Dream party" received more votes overall. The distribution of votes from $40\%$ to  $90\%$ exceeds all votes received by opposition parties, therefore it is reasonable to state that the ``Georgian Dream Party" accrued high percentages in unusually many precincts. In Fig.\ref{fig2} we show a scatter plot of the the turnout percentage (depicted on the X axis) and the ``Georgian dream" share of the turnout (depicted on the Y axis). Each circle represents an election precinct and its size is defined by the total turnout of the precinct-- the more the turnout the bigger the size of the circle-- and its color represents the density of the precinct. This graph, which was constructed using the official data published by the Central Election Commission shows that there are four precincts (indicated by the red ellipse on the graph) in which more people showed up than were registered, resulting in a turnout percentage of more than  $100\%$. In these four precincts "The Georgian Dream Party" had more than $80\%$ of the votes. There are a couple of precincts (indicated by the green ellipse on the graph) where there is no information about the total number of registered voters given by the ``CEC" and at all of these precincts, ``The Georgian Dream Party" received more than $70\%$ of the votes. Most precincts are localized on the graph but a large number of election precincts are spread around the graph, in which "The Georgian Dream Party" received more than $70\%$ of the votes in the vast majority of precincts.
At first glance, the graphs suggest that there is a possibility that votes were manipulated, however, it is difficult to conclude anything with these graphs alone. Due to this, we decided to construct a program to compare the computational results and the official data to investigate possible manipulation.

\section{Computer simulation}\label{sec3}

Our aim was to investigate the distribution of votes across precincts in the case of no election manipulation under the assumption that ``The Georgian Dream Party" received approximately the same number of overall votes as they did in the election: $54\%$. To explore this, we designed a computer program replicating the voting process and ensuring that the party achieved the same $54\%$ result overall. By comparing this simulated distribution to the actual election results, we can identify potential anomalies or irregularities.

Even if the opinions are overly polarized within a single electoral district, the party’s performance should remain relatively consistent across different precincts. This is because voters are assigned to precincts randomly, often based on their residential streets. Consequently, even in districts with sharply divided opinions, all precincts should contain a representative mix of voters, leading to similar (not drastically different) results for the party across precincts. We assigned each voter a probability to vote for ``The Georgian Dream Party" based on the official percentage of the party in the electoral district with a ``Gaussian Distribution" with a predefined mean, which is the official percentage. We chose the Gaussian distribution as it is the most common model to describe statistical processes in real life~\cite{p}. However, there was the small problem of choosing the standard deviation $\sigma$, which we calculated as follows:

\[
\sigma = \sqrt{\frac{1}{N} \sum_{i=1}^{N} (x_i - \mu)^2}
\]

where $N$ is the total number of election precincts in the district, $x_i$ represents the percentage of votes for the Georgian Dream Party in the i'th election precinct in the district and $\mu$ is the mean probability for the district. We considered two different approaches to this problem, resulting in two different simulations.

In the first approach, $\sigma$ was the same for each electoral district. In this case, we did not need to choose anything manually because the requirement that the "Georgian Dream Party" must achieve $54\%$ overall fixed the parameter $\sigma$ within a very small range $0<\sigma<0.1$. For other values of $\sigma$, the party's final result would differ from the official results. If it differed, then we didn't include it in our analysis, since the main task was to find what the distribution should have been to obtain the official results without any manipulations from the government(``The Georgian Dream Party"). Interestingly, this range matched the official data from "CEC".

In the second approach, we calculated the parameter $\sigma$ for each electoral district based on the real official data. In this case, the range of $\sigma$ was also very small and matched the range $0<\sigma<0.1$ obtained in the first approach. However, when using this individual $\sigma$ approach for each electoral district, it was impossible to achieve the overall $54\%$ result. Instead, the results consistently ranged from $51\%$ to $52\%$ for ``The Georgian Dream Party." This discrepancy is a strong indication of manipulation in the election process itself, as it means that there were drastically different results in the same electoral district but in different election precincts. However, it did not provide information about the scale and number of the manipulated votes. Therefore, we used the first approach and constructed the code accordingly. We conducted simulations for 10 different values of $\sigma$ within the range $0 < \sigma < 0.1$. For each $\sigma$, the code was executed $250$ times, resulting in a total of $2,500$ simulated elections. From these, we retained only the outcomes where ``The Georgian Dream Party" secured between $53\%$ and $55\%$ of the votes. Although we could have narrowed this range, we opted for a slightly broader interval to allow greater flexibility. This filtering process excluded approximately $70\%$ of the results, leaving us with $803$ simulated elections for further analysis.

\section{Results}\label{secn}

\begin{figure}
\resizebox{\hsize}{!}{\includegraphics[width=10cm]{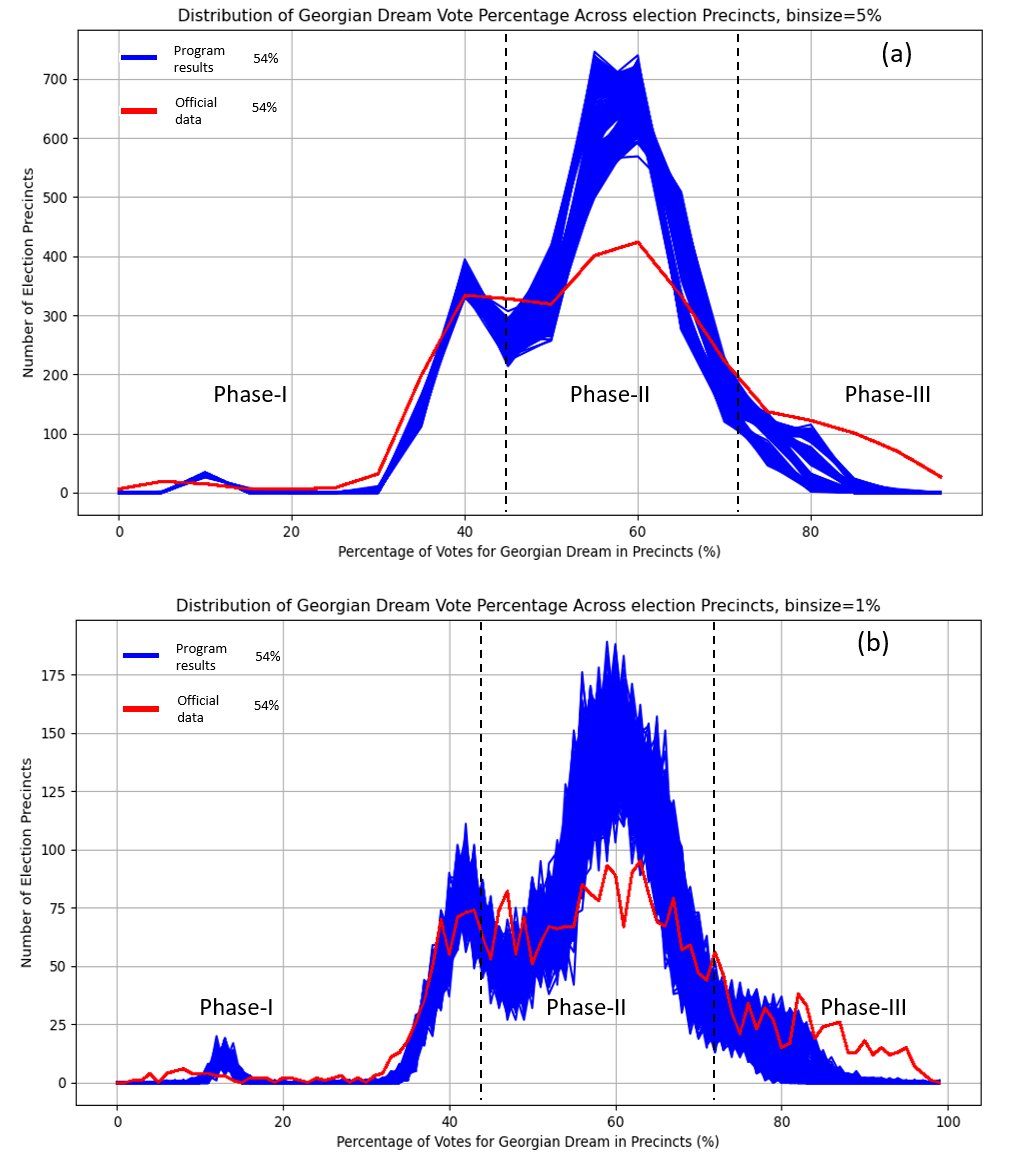}}
\caption{Panel (a) shows the distribution of precincts based on votes received by the ``Georgian Dream Party" with a bin size of $5\%$. Blue lines represent results from the $803$ simulated elections, red line represents the official results from the election. Panel (b) presents the same distribution with a finer bin size of $1\%$.
} \label{fig3}
\end{figure}

We present the results of this analysis in Figure \ref{fig3}. In red we show a histogram of the percentage of votes received by ``The Georgian Dream Party" across precincts. The x-axis represents the percentage of votes obtained in each precinct, divided into bins of size $5\%$ in panel (a) and bins of size $1\%$ in panel (b). The y-axis corresponds to the number of precincts falling within each percentage bin. In blue we show the resulting histograms from our $803$ simulated elections. This approach allowed for a clear representation of how vote percentages are distributed across all precincts, highlighting patterns and potential irregularities. These graphs are histograms, but they are represented using points and lines for improved clarity and a more intuitive view. We divided these graphs into three phases. From around $0\%$ to $45\%$  of the votes, which we refer to here as Phase I, the simulations and the official election results aligned fairly well. We found that a large majority of the precincts that fall within this Phase are urban precincts. Phases II and III correspond to rural areas, where a clear distinction can be observed. We conclude that, without any manipulation, the party should have achieved a significantly higher number of precincts with approximately $60\%$ of the votes. However, this did not occur. Instead, the vast majority of votes "lost" in Phase II appear to have been compensated for in Phase III. Based on this distinction, we calculated the number of manipulated votes by calculating the difference between our computational results and real data in Phase II which was mostly compensated by Phase III. The results indicate that the minimum number of manipulated votes is approximately $140,000$, the maximum is $200,000$, and the most probable number is around $175,000$ votes.

Panel (b) represents a similar graph but provides a more microscopic view of the elections, as the percentage range (bin size) is set to $1\%$. While the overall tendency of the distinction and the graph's shape remains consistent with panel (a), the range of manipulated votes shifts to between $90,000$ and $245,000$. However, the most probable number of manipulated votes remains unchanged. In panel (b), in Phase I, a new sharp peak in the generated results appears. Upon analyzing these precincts, we observed that all of them are located outside of Georgia, which we treated as a single electoral district in the code. While this does not perfectly reflect reality, we made this decision due to the low percentage of votes for ``The Georgian Dream Party" outside of Georgia and the relatively small total number of voters in these precincts, approximately $30,000$.

\section{Discussion}\label{sec4}
As described in the previous section, the number of manipulated votes is not small, in this case, it's quite large and changes election results significantly. We've found that there is evidence that the manipulated votes ranged from $ 140,000$ to $200,000$ votes for larger bin size and from $90,000$ to $245,000$ for smaller bin size, where in both cases the highest probability of falsification was observed at $175,000$. We assume that from the $175,000$ manipulated votes, a certain number of voters wouldn't go to the polls at all, and the remaining number would vote for the opposition if not for bribery, coercion, etc. We have to bear in mind that stolen votes damaged the opposition twice which means that it would have a bigger effect on the elections than we determined from our code.

\section{Conclusions}
In this paper, we investigated possible election fraud in Georgia. We collected official data from the CEC website from which we constructed distribution graphs. We then built a density graph that included the turnout information which shows us anomalous behaviors, such as some election precincts having more people show up than were registered, some precincts with no information about the registered number of voters, and all of the suspicious precincts having a percentage of $70\%$ or higher for ``The Georgian Dream Party". We then wrote code based on official data and showed a conflict between the CEC (Central Election Commission) data and results from our simulations. In our first approach, we calculated $\sigma$ in the range $0<\sigma<0.1$ for every electoral district separately, however with this approach ``the Georgian Dream Party" never got close to $54\%$ and our computational results remained in the range between $51\%$ to $52\%$ further providing strong evidence of a manipulated election. In our second approach where assumed a $\sigma$ that was the same in every district, the results where the winning party gets $54\%$ give rise to a mismatch of about $140,000$ to $200,000$ votes for the $5\%$ bin size and about $90,000$ to $245,000$ votes the for $1\%$ bin size. In both cases, the most probable number of manipulated votes is approximately $175,000$ which would have changed the election results dramatically. We saw that the election outcome would be drastically different if not for interference and manipulation of the votes and concluded that the election was rigged.

\bmhead{Acknowledgments}

We would like to thank, Sophie Dvali, Luka Rapava, Levan Tserediani, and Giorgi Tserediani for their support and valuable discussions.

\bmhead{Data Availability Statement}: The datasets generated during analyses and all the codes can be seen in a GitHub Repository https://github.com/RocinantEL/Rigged-Elections-of-Georgia-2024.

\end{document}